# Lasing action in single subwavelength particles supporting supercavity modes

Vasilii Mylnikov[1,2], Son Tung Ha[1], Zhenying Pan[1], Vytautas Valuckas[1], Ramón Paniagua-Domínguez[1], Hilmi Volkan Demir[2,3], Arseniy I. Kuznetsov[1#]

[1] Institute of Materials Research and Engineering, A*STAR (Agency for Science, Technology and Research), 2 Fusionopolis Way, #08-03, Innovis 138634, Singapore

[2] LUMINOUS! Center of Excellence for Semiconductor Lighting and Displays, The Photonics Institute, School of Electrical and Electronic Engineering, School of Physical and Mathematical Sciences, Nanyang Technological University, 50 Nanyang Avenue, Singapore 639798

[3] Bilkent University UNAM - Institute of Nanotechnology and Materials Science, Department of Electrical and Electronic Engineering, Department of Physics, Bilkent University, Ankara, Turkey 06800

[#] Corresponding author, email: arseniy_kuznetsov@imre.a-star.edu.sg

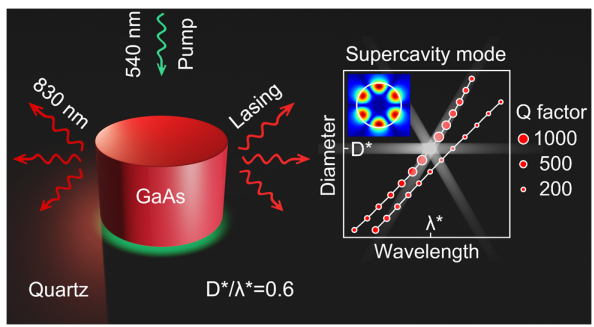

**ABSTRACT**

On-chip light sources are critical for the realization of fully integrated photonic circuitry. So far, semiconductor miniaturized lasers have been mainly limited to sizes on the order of a few microns. Further reduction of sizes is challenging fundamentally due to the associated radiative losses. While using plasmonic metals helps to reduce radiative losses and sizes, they also introduce Ohmic losses hindering real improvements. In this work, we show that, making use of quasi-bound states in the continuum, or supercavity modes, we circumvent these fundamental issues and realize the smallest purely semiconductor nanolaser thus far. Here, the nanolaser structure is based on a single semiconductor nanocylinder that intentionally takes advantage of the destructive interference between two supported optical modes, namely Fabry-Perot and Mie modes, to obtain a significant enhancement in the quality factor of the cavity. We experimentally demonstrate the concept and obtain optically pumped lasing action using GaAs at cryogenic temperatures. The optimal nanocylinder size is as small as 500 nm in diameter and only 330 nm in height with a lasing wavelength around 825 nm, corresponding to a size-to-wavelength ratio around 0.6. The obtained results pave the way for the development of smaller on-chip light sources free of Ohmic losses, which may find applications in the future photonic circuits.

In the last decade, resonant dielectric nanoantennas have emerged as a promising platform for nanophotonic applications. Low Ohmic losses and CMOS compatibility, accompanying small form factors, are the key advantages of these systems that make them particularly attractive for industrial applications as compared to prior designs based on plasmonic metals[1]. In recent years, moreover, significant efforts are being made in combining dielectric nanoantenna concepts with active material platforms[2], to extend this promising approach towards efficient directional light sources. By now, the use of dielectric nanoantennas in emission processes has been mainly focused on enhancing fluorescence from molecules[3-8], enhancing the Raman signal[9-11] and enhancing photoluminescence (PL) from the nanoantenna material itself[12-15]. Due to the limited quality factors (Q factors) of low-order modes excited in dielectric nanoantennas, lasing has not been yet demonstrated for particle sizes significantly smaller than the wavelength, with the recent exceptions of micrometre-sized systems in which sub-wavelength nanoantennas are arranged in lattices[16,17]. Hence, besides strongly dissipative systems involving metals[2,18], on-chip lasers have been mostly restricted to spatially extended systems with sizes on the order of a few microns.

The smallest fully dielectric lasers were experimentally demonstrated using suspended disk structures[19,20]. High-order Mie-type modes (i.e., whispering gallery modes) were used in those designs to achieve lasing. In this regard, the smallest nanolaser reported consisted of a disk on top of a pedestal with a diameter of 627 nm, operating in the 4$^{th}$ order Mie resonance and emitting at a wavelength of 870 nm, which corresponds to a size/wavelength ratio $\zeta=0.72$[20]. Very recently, one of the smallest, fully dielectric, single-particle-based laser has been demonstrated using perovskites in a nanocube configuration[21]. This also uses the 4$^{th}$ order Mie type mode and has a slightly larger size/wavelength ratio of $\zeta=0.79$ but operates at room temperature. Further reduction of size (e.g., in disk diameter) is challenging due to two different reasons. First, using smaller resonators while keeping the same operating wavelength implies necessarily the use of lower-order modes. Those present higher radiative losses, which leads to a drop in the Q factor. Second, smaller particles have larger surface-to-volume ratios, which increases the effect of surface recombination and roughness, resulting in higher losses that make the gain-loss compensation even more difficult. While the second is more difficult to overcome, without improving the fabrication processes and/or including some surface passivation methods, the first still allows some improvement from the design perspective. Thus, new approaches need to be implemented for improving the quality of the cavities in order to achieve lasing in smaller resonators.

A recent promising approach to boost the Q factor of a system is engineering it to support a special type of resonant mode called bound state in the continuum (BIC). BICs are non-radiative states, thus perfectly spatially confined, which nevertheless coexist with the continuum of radiative states (i.e., despite the presence of available radiation channels)[22]. BICs are a general wave phenomenon, first reported in electronic systems and subsequently extended to photonic ones[22], and are true eigenmodes of the system (in fact, often referred to as embedded eigenstates). Strictly speaking, for any system comprising regular materials with ε≠0, µ≠0, ε≠±∞ or µ≠∞ BICs can only be supported if the system is infinite, and their Q factors are infinite[22]. Parameters of a finite system, however, can be tuned close to the condition of BICs in the corresponding infinite system[23]. The obtained quasi-BICs are often referred in literature to as supercavity modes and possess finite but high Q factors[23]. This makes supercavity modes or quasi BICs useful for many photonics applications,[22,24] including, for example, sensing[25,26] and high-harmonic generation[27-30].

Regarding the use of BIC for lasing, one of the first quasi-BIC lasers was demonstrated in 1985. It supported a symmetry protected BIC by supressing radiation into the transverse directions in a 1D distributed feedback laser cavity[31]. After that, similar quasi-BIC principles were demonstrated in many photonic crystal surface-emitting laser devices[22]. In recent years, new quasi-BIC-based lasing results were shown for on-chip systems based on nanoparticle arrays[16,17]. Going beyond the BIC supported by spatially extended systems towards single particle configurations, it should be mentioned that a spatially confined system can only support true BIC if it contains some region in which the permittivity tends to zero[32,33]. Nevertheless, several recent reports have theoretically predicted that it should, still, be possible to obtain, if not a true BIC, but a quasi-BIC or supercavity mode in such a simple system as a single dielectric nanocylinder[28,34,35]. This quasi-BIC has a Friedrich–Wintgen nature and allows some degree of radiation to escape the system. Thus, by reciprocity, it is also possible to excite it by external sources. A recent experimental work has, indeed, demonstrated the excitation of these quasi-BICs in a single cylinder at GHz[36] and optical[30] frequencies. However, no experimental results of lasing in supercavity regime for a single semiconductor nanoparticle have been showed so far. In this work, elaborating on the BIC concept, we report the first single dielectric nanoparticle laser working in a supercavity regime. This also enables the realization of the first fully dielectric nanolaser working in the third azimuthal order mode for a particle on a substrate, with a record size/wavelength ratio as low as ζ=0.6.

We design our cylindrical nanoresonator to support a quasi-BIC that originates from the strong interaction of two resonant modes: a Mie-like mode (which strongly depends on the diameter of the cylinder and is largely insensitive to its height) and a Fabry-Perot (FP)-like mode (which is formed by the multiple reflections of the guided mode supported by the cylinder at its top and bottom interfaces, therefore highly sensitive to the cylinder height[34]). The interaction between these two resonant modes manifests itself as an avoided crossing between them in the scattering spectrum of the cylinder. Along with the avoided crossing, one of the modes exhibits linewidth narrowing (increase in the Q factor) along with a decrease in the light scattering efficiency in the vicinity of the optimal geometrical parameters leading to the quasi-BIC. Thus, there are three main signatures of the emergence of this type of supercavity modes: anticrossing between Mie-like and FP-like modes, increase in the Q factor of one of them and decrease in its scattering efficiency. The use of a supercavity mode reduces the radiative losses of the system (increasing its Q factor several times with respect to a regular Mie mode of the same order[34]) enabling gain/loss compensation in nanoparticles of a smaller size.

We focus our attention on obtaining lasing from a supercavity mode with azimuthal order $i$ = 3. In particular, we study a single GaAs cylinder standing on a quartz substrate. Due to the relatively low gain of GaAs at room temperature, which is further reduced by the large surface recombination present in small particles, lasing experiments are performed at the liquid nitrogen temperature (i.e., 77 K) to increase the quantum yield of GaAs (note that no surface passivation or any chemical treatment is applied to the nanostructures). When tuning the temperature from 300 to 77 K, the emission peak of GaAs shifts from 870 to 830 nm[16]. Thus, in this work, the geometrical parameters of the cylinder (diameter D and height H) are optimized to achieve high Q factors at λ=830 nm.

To illustrate the physical mechanism behind the supercavity mode formation in a simple way, let us first consider a cylinder with a constant relative permittivity ε=13 (corresponding, approximately, to GaAs at 830 nm wavelength and 77 K) in vacuum. We perform two types of three-dimensional electromagnetic simulations – scattering simulations and Q factor calculations (illustrated in Figures 1A and 1B respectively). For the scattering simulations we study illumination of the cylinder by a transverse electric (TE) wave. TE wave is defined as the one having incident magnetic field polarized along the axis of the cylinder and the electric field and incident wave vector perpendicular to it as shown in Figure 1A. In the Q factor simulations, we excite TE modes by a point-like dipole highlighted in Figure 1B (see Methods section for more details). We compute the scattering efficiency and the eigenmodes supported by this cylinder as a function of its geometrical parameters using full wave simulations based on the finite-difference time-domain method (see Methods section for the detailed description). The results are shown in Figure 1C. There, we plot the scattering efficiency as a function of the dimensionless parameters $x_H = \pi H/\lambda$ (i.e., the size parameter, where the height of the cylinder is used as a characteristic dimension) and $\rho = D/H$ (the cylinder aspect ratio). Overlaid with the scattering map, we plot the quasi-TE eigenmodes (quasi-normal modes) corresponding to the azimuthal order $i = 3$ by a set of solid circles, their area being directly proportional to the Q factors of the modes. As seen in the plot, two modes with this azimuthal order coexist in the range of size parameters studied (denoted as 3 and 3'). More important is the clear indication of their strong coupling, evidenced through their avoided spectral crossing. Together with this, a dramatic increase in the Q factor, leading to a maximum, and a noticeable reduction of the associated scattering efficiency of mode 3 are observed near the anti-crossing region, indicating that a supercavity mode has been formed. The optimum values of dimensionless parameters for this 3$^{rd}$ azimuthal order supercavity mode are $x_H \sim 1.30$ and $\rho \sim 1.47$. For an operational wavelength $\lambda = 830$ nm, these correspond to a cylinder with height $H \sim 344$ nm and diameter $D \sim 504$ nm. The corresponding Q factor reaches a maximum value of Q = 970. It should be noted that it has been theoretically shown[34] that an infinite permittivity is formally equivalent to a spatially infinite system in which an ideal BIC may exist. Practically, this means that the higher the permittivity is, the closer to an ideal BIC the supercavity mode is. This explains why in the case of moderate permittivity of GaAs, the supercavity formation leads to a somehow moderate 5-fold increase in the Q factor. This, nevertheless, turns out to be crucial to reduce the lasing threshold for the 3$^{rd}$ azimuthal order mode, as will be shown in the experiment below.

To complete the characterization of the supercavity mode, the near fields and the far-field radiation pattern near the supercavity regime are depicted in Figures 1D-G. From the near-field maps (Figures 1D, 1E), the TE$_{302}$ character of the mode becomes apparent (in TE$_{ijl}$ notation *i*, *j*, and *l* are the azimuthal, radial and axial indices of the mode, respectively). On the other hand, the far-field pattern (Figure 1F) shows that, along with the regular in-plane lobes, which are common for pure Mie modes, out-of-plane lobes are present as well. This can be explained by the fact that supercavity mode is a hybrid FP-Mie mode and, thus, it cannot be identified as a pure FP mode or a pure Mie mode. A detailed analysis of the evolution of the near-field distribution around the supercavity regime reveals that mode 3 has indeed three different regimes, evolving from a Mie-like one (Figure 1C at point α) to the Supercavity one (Figure 1C at point β) and finally to a Fabry-Perot-like one (Figure 1C at point γ).

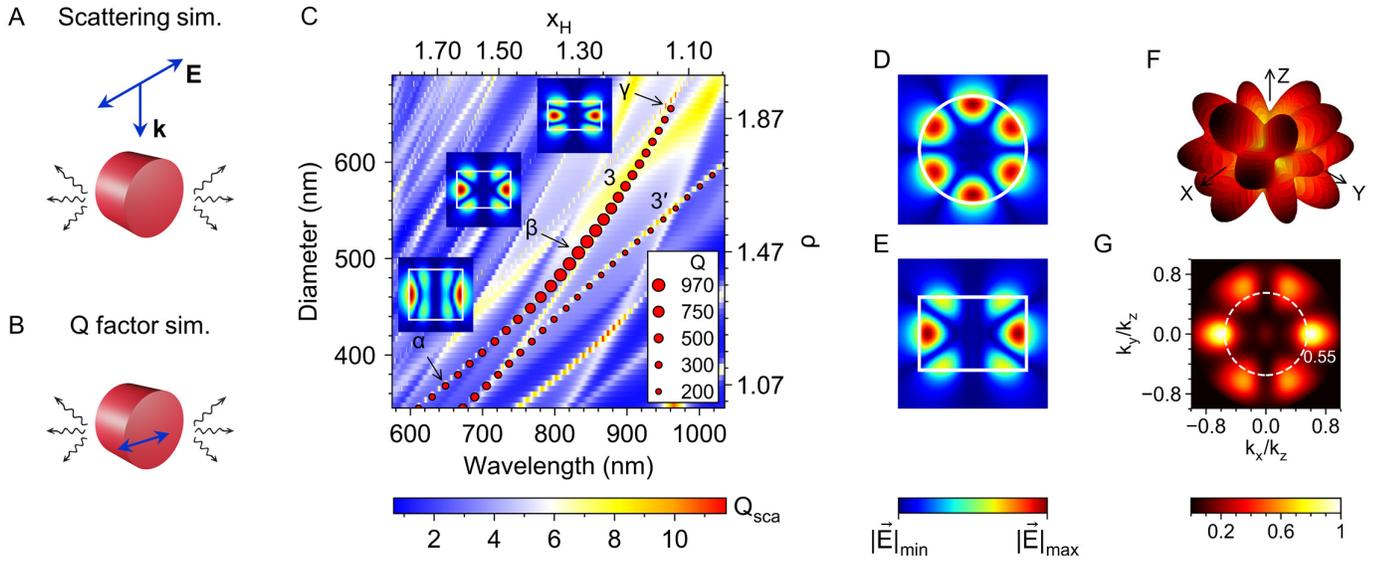

**Figure 1.** Simulated characteristics of a single GaAs nanocylinder (ε=13) in vacuum. (A, B) Illustration of the dielectric cylinder in vacuum corresponding to scattering simulations (A) and Q factor (near field, far field) simulations (B). (C) Scattering efficiency map Q$_{sca}$ as a function of the dimensionless parameters x$_H$ = πH/λ (the size parameter) and ρ = D/H (the cylinder aspect ratio). The quasi-TE modes with azimuthal order *i = 3* (labelled 3 and 3') supported by the cylinder are also shown by red solid circles, their area being directly proportional to the value of the Q factor. Only these two modes are shown in order to highlight their anti-crossing and Q factor evolution as a function of the diameter and wavelength (the Q factor reaches its maximum value for the optimal parameters indicated by a point β). The alternative x- and y-axis correspond to the wavelength and diameter obtained when the height of the cylinder is fixed at H=345 nm. The insets

show the electric near-field maps in the YZ plane passing through the centre of the cylinder for different regimes of the mode 3. (D, E) Simulated near fields of the supercavity mode at the optimal geometrical parameters in the XY (D) and YZ (E) planes of the cylinder, respectively. White lines in (D, E) show the cylinder boundaries. (F) Simulated 3D far-field radiation pattern corresponding to the point β. (G) 2D projection of the simulated far-field radiation pattern corresponding to the point β. The white dashed circle shows the range of wave-vectors collected by a 0.55 NA objective (similar to the one used to collect the emission in the lasing experiment).

To experimentally demonstrate lasing with the analysed 3$^{rd}$ order supercavity mode, GaAs nanocylinders standing on a quartz substrate are fabricated using the method reported in Ref[16]. First, a GaAs film grown on top of a commercially available GaAs wafer with a sacrificial layer of AlAs is transferred to a quartz substrate using the technique known as epitaxial lift-off[37] followed by a direct bonding process developed in-house using a wafer bonding system. The transferred film is then etched to a desired thickness using inductively coupled plasma (ICP) method. The nanocylinder pattern is then defined using standard e-beam lithography with hydrogen silsesquioxane (HSQ) as a negative resist. The final structure is formed by an additional dry etching step (see Methods for a more detailed information). SEM image of one of the fabricated nanostructures is shown in the inset of Figure 2B. Note that, despite their low index, the presence of the quartz substrate and the remaining capping HSQ layer affects the spectral position and the Q factor of the supercavity mode. Thus, we reoptimize the parameters of the cylinder to account for them, obtaining a maximum Q factor of 280 for optimum values of the dimensionless parameters $x_H$=1.29 and $\rho$=1.47, corresponding to the cylinder height H=340 nm and diameter D=500 nm at for λ=830 nm. The height of the GaAs cylinders obtained in the fabricated samples is close to the targeted one, around 330 nm (measured using reflectometer and tilted SEM images). The sample is then cooled down to 77 K using liquid nitrogen for further optical characterization (see Methods for details).

Before characterizing the fabricated nanostructures, we first investigate the absorption and PL spectra of GaAs films with the thickness of 350 nm. The absorption spectrum, recorded at 77 K, is presented in Supplementary Figure S1. The spectrum clearly shows the absorption edge at ~ 827 nm, in good correspondence with values reported in literature for GaAs at this temperature[16]. The narrow absorption peak at ~ 830 nm can be attributed to the excitonic resonance of the material. For the spontaneous emission spectrum, large square areas (of 50 μm x 50 μm) patterned in the film using EBL, are analysed. We do so to characterize, together, the emission of GaAs and that of HSQ resist, used to pattern the nanostructures in our experiments and known to have non-negligible fluorescence when treated with e-beam and high temperatures[38,39]. Indeed, for GaAs film with HSQ resist on top, the measured photoluminescence (Figure 2A) has a broad emission spectrum from 700 to 850 nm (a long pass filter is used to cut the pump at 700 nm), beyond that for GaAs alone which has a relatively narrow PL spectrum from 800 to 850 nm. The above-mentioned broad spectrum is attributed to the photoluminescence of GaAs combined with that of silicon nanocrystals formed in the silica matrix of cured HSQ (whose PL can range from 650 to 800 nm[39]), which, in our case, most probably arises due to the high energy electron beam exposure, in a chemical reaction analogous to that happening during thermal annealing[40]. Thus, in our experiments, the photoluminescence of both Si nanocrystals in cured HSQ[41] and GaAs itself can provide gain for our nanolasers with different resonance wavelengths.

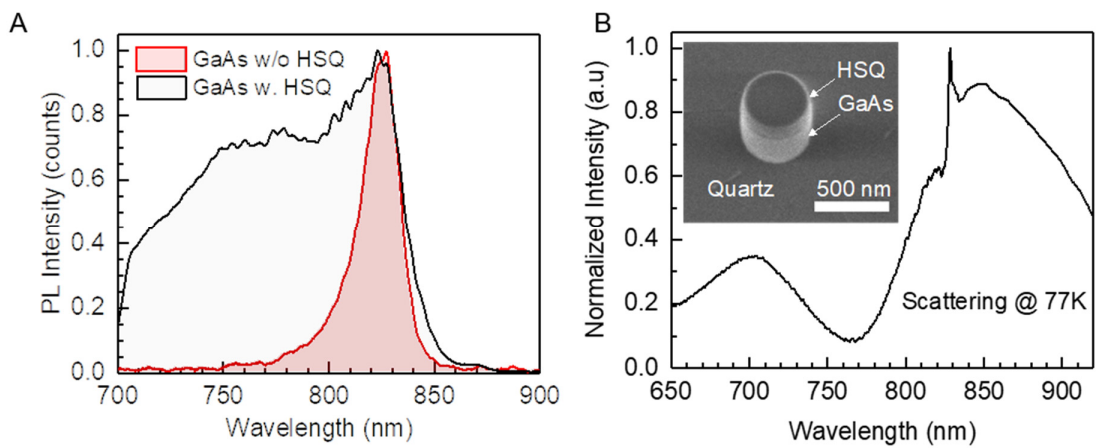

**Figure 2.** (A) Normalised photoluminescence spectrum measured from GaAs film without HSQ (red) and with 150 nm layer of HSQ on top (black). (B) Normalised experimental dark-field scattering spectrum of a single GaAs nanocylinder with parameters D=500 nm and H=330 nm. Inset: exemplary SEM image of the fabricated nanostructure.

With this information in mind, we now move to the optical characterization of the nanocylinders at cryogenic temperatures, first studying their scattering spectra, to identify the formation of the supercavity mode, and then their emission under optical pumping. Figure 2B shows the characteristic dark-field scattering spectrum of a single GaAs nanocylinder with D ~ 500 nm and H ~ 330 nm (as measured using SEM). From our theoretical analysis, we expect the cylinder to support the supercavity mode at around 830 nm for these dimensions. Indeed, one can observe the emergence of a very narrow peak in the experimental spectrum at ~ 828 nm, in very good agreement with what is expected from the simulations, thus confirming the presence of the quasi-BIC mode. To demonstrate lasing action, we optically pump the nanocylinders using a femtosecond laser (with a wavelength of 530 nm, 200 fs pulse duration and 20 KHz repetition rate) and collect the PL using a micro-spectrometer setup (see detailed information in Methods section). Figure 3A shows the PL spectrum evolution of the GaAs nanocylinder with diameter D ~ 500 nm and height H ~ 330 nm, supporting the supercavity mode, when pumped under different laser fluences. It can be seen that when the pumping fluence is larger than 260 µJ/cm$^2$, a pronounced, narrow peak appears at ~ 825 nm. By plotting the integrated emission intensity versus the pumping fluence in log-log scale (Figure 3B) one can clearly see the "S" shape that indicates the transition from spontaneous emission (SE) to amplified spontaneous emission (ASE) and finally to lasing, with an ASE threshold of ~ 255 µJ/cm$^2$. The PL peak position and full-width-at-half-maximum (FWHM) as a function of the pumping fluence, extracted from the spectra, are shown in Figure 3C. Based on the FWHM, we observe a maximum Q factor of 235 in the lasing regime (in good agreement with the theoretically calculated value of 280, as discussed above). From the analysis of the lasing peak position, a clear blue-shift is observed when the pumping fluence is increased, which can be attributed to the Burstein-Moss effect associated to the band-filling[42,43]. In addition to spectral measurements, emission directivity patterns are also obtained using the back-focal plane imaging technique. The results can be found in the Supplementary Information (Figures S2A and S2B).

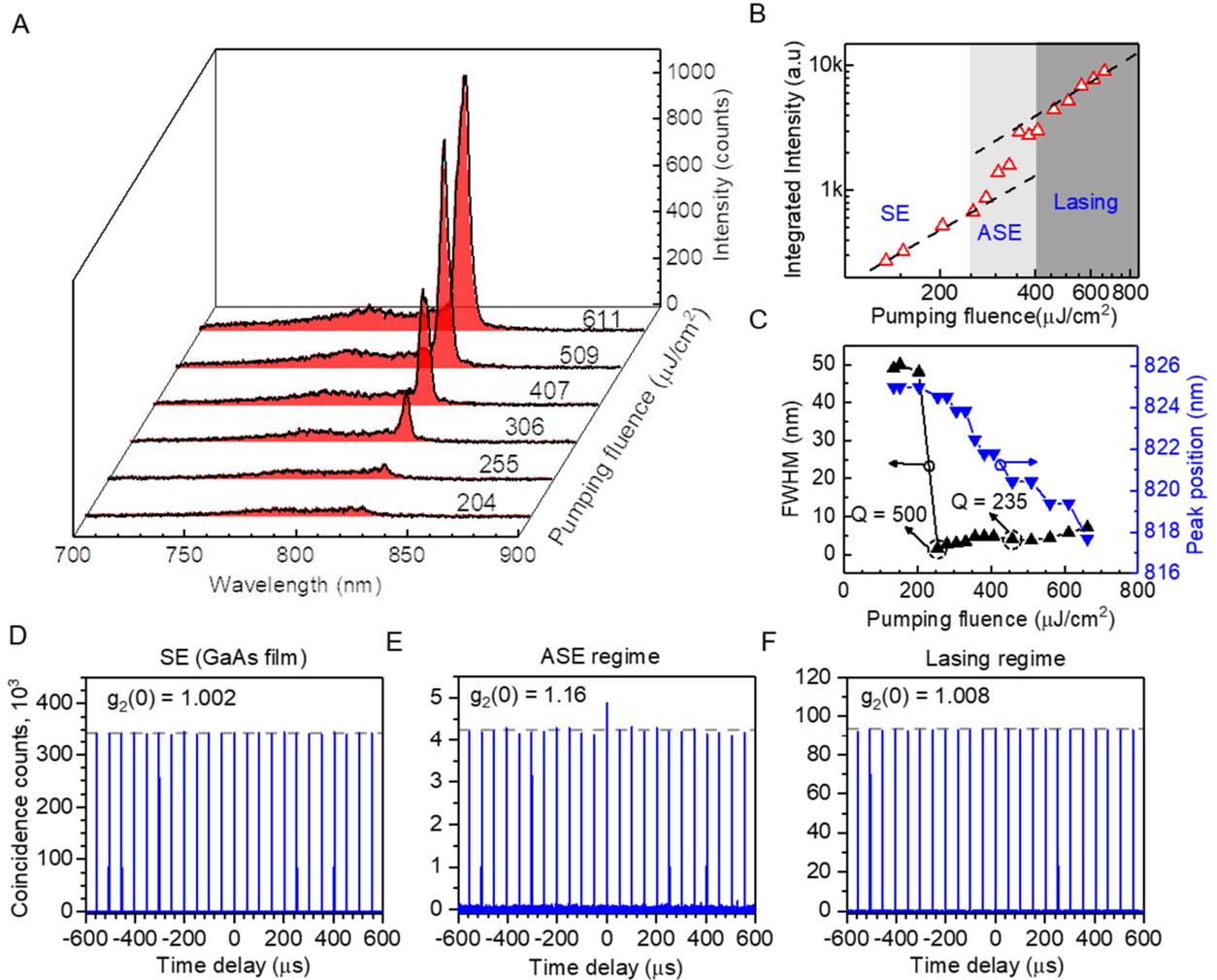

**Figure 3.** Lasing action in a GaAs nanocylinder. (A) Evolution of the emission spectrum of the GaAs nanocylinder with diameter D=500 and height H=330 nm at different pumping fluences. (B) Log-log plot of output emission intensity versus pumping laser fluence showing the transitions from spontaneous emission (SE) to amplified spontaneous emission (ASE) and finally to lasing. (C) Peak position and FWHM of the lasing peak extracted from the emission spectra. (D) Intensity auto-correlation histogram for 50 μm X 50 μm GaAs square at the pumping laser fluence F=764 μJ/cm² and for the single GaAs nanocylinder of diameter 500 nm and height 330 nm for the pumping fluences (E) F=382 μJ/cm² and (F) F=433 μJ/cm².

To further confirm the coherent nature of the lasing emission, autocorrelation measurements using a Hanbury–Brown–Twiss set-up[16,44] are performed. Figures 3D-F show coincidence counts at different time delays for three different regimes: SE (Figure 3D), ASE (Figure 3E), and lasing (Figure 3F). The normalised second-order correlation function at zero time delay $g_2(0)$ is defined as the ratio between coincidence count number of the pulse at zero time delay and an averaged count number over 22 other neighbouring pulses. It is noted that the auto-correlation measurement for SE (Figure 3D) is performed on relatively large patterned GaAs squares (of 50 μm X 50 μm). This is to ensure that an adequate amount of signal can be collected and analysed. It, however, does not change the nature of the SE from the materials composing the nanolaser. Due to their larger size, the photoluminescence intensity of these squares is several times greater than that of the single GaAs nanocylinder used to measure $g_2(0)$ in the ASE and lasing regimes, which explains why the coincidence counts in Figure 3D is substantially higher than those of Figures 3E and 3F. The $g_2(0)$ for the SE case is calculated to be 1.002, which is significantly smaller than the theoretical value for an ideal thermal light source (i.e., $g_2(0) = 2$). This is because the coherence time of the SE is much lower than the detection limit of our experimental setup (~ 81 ps), in accordance with similar observations reported in earlier studies[16,45]. Nevertheless, we are able to measure a clear decrease in $g_2(0)$ upon transitioning between the ASE and lasing regimes, as shown in Figure 3E and 3F, the $g_2(0)$ reducing from 1.16 (for ASE regime, F=382 μJ/cm$^2$) to 1.008 (for lasing regime, F=433 μJ/cm$^2$) and, thus, clearly indicating the coherent behaviour of the nanolaser.

We now prove that lasing is indeed obtained for the third azimuthal order supercavity mode by studying how a slight detuning of the geometrical parameters from the optimal ones affects the performance of the system. For that, lasing and dark-field scattering measurements are carried out for cylinders of the same height, namely H=330 nm, but with diameters above and below the optimum one. In particular, Figure 4A shows the dark-field scattering spectra of GaAs cylinders with diameter ranging from 420 to 653 nm with an increment of ~ 5 nm. The mode highlighted by green dashed line in Figure 4A corresponds to the 3$^{rd}$ azimuthal order TE mode from the simulation (see Figure 4B). Higher- and lower-order modes can also be seen from the scattering data in Figure 4A.

For a better visualization of the formation of the supercavity mode, the scattering spectra for several selected radii around the optimum one are depicted in Supplementary Figure S3A. It should be noted that, due to the presence of high absorption at wavelengths below the excitonic resonance, it is not possible to observe the full evolution of the scattering peak associated to the formation of the supercavity mode (from broad to narrow and back to broad) but only the narrowing of the resonance. Thus, we explore an alternative experimental method, relying on the analysis of lasing threshold, to prove that the 3$^{rd}$ azimuthal order TE mode used for lasing indeed undergoes the formation of the supercavity regime, as explained below.

Figure 4B shows the theoretically obtained scattering peak locations of the modes in the span of parameters λ and D of our interest. In the case of the 3$^{rd}$ azimuthal order TE mode, the corresponding Q factors are also plotted by red circles, so that the supercavity regime can be readily identified. Superimposed in the same plot, the green triangles depict the experimental lasing peak positions obtained for various disk diameters. As can be seen, there is a good agreement between the simulated scattering peak positions and the lasing peaks observed in the experiment. The discrepancy between theory and experiment is minimal and lays within the error of diameters determined by SEM and the possible slight deviations in the material parameters. Importantly, we can achieve lasing for the 3$^{rd}$ azimuthal order TE (3 TE) mode for a relatively wide range of cylinder diameters due to the low divergence of quality factor near the optimal value. This gives us flexibility to tune the emission wavelength of the nanolaser within the gain spectrum of the active materials involved (note that, in our case, these refer not only to GaAs, but also to the Si nanocrystals obtained inside HSQ resist after the e-beam exposure). Besides the mode 3 TE, lasing is also obtained for larger nanocylinders supporting higher azimuthal order modes (up to 5$^{th}$ order TM mode as shown in Figure 4B). It is important to note that, for nanocylinders with diameters close to the optimum (D ~ 500 nm), the 3$^{rd}$ order mode of interest (3 TE undergoing the supercavity regime) and the 4$^{th}$ order Mie mode (4 TM) are spectrally close (~ 20 nm away from each other, according to the theoretically computed scattering spectra, as can be seen in Figure 4B). Thus, it is crucial to observe both the 3$^{rd}$ and 4$^{th}$ order modes in the lasing experiments to unequivocally prove that the mode 3 TE is identified correctly. The detailed view of the lasing spectra of both these modes when they lase simultaneously is shown in Supplementary Figure S3B. Supplementary Figure S4 also shows additional information on the modes 3 TE and 4 TM obtained from the simulations. Moreover, as can be seen in Figure 4B, the lasing peak positions for all modes (3$^{rd}$, 4$^{th}$ and higher azimuthal orders) match well with the theory, further supporting the correct identification of the supercavity mode.

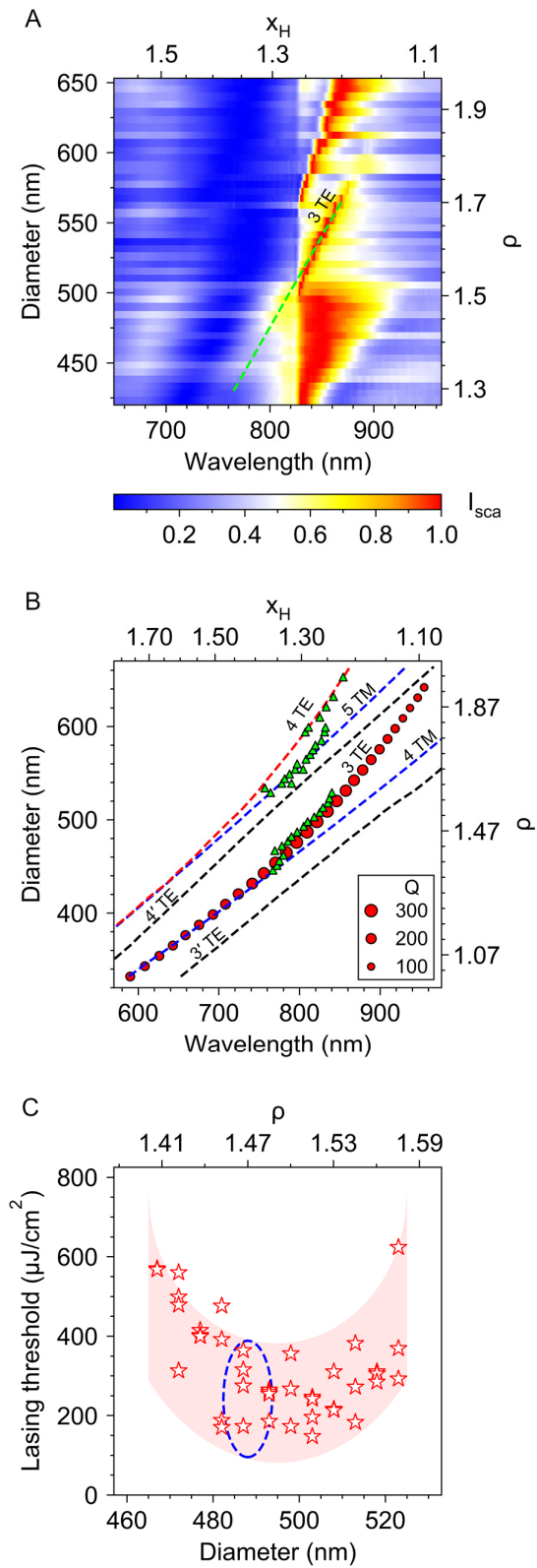

**Figure 4. Lasing characteristics of the supercavity mode in the GaAs nanolaser.** (A) Normalized experimental dark-field scattering spectra of GaAs nanocylinders with fixed height H=330 nm and varying diameter D = 420 – 653 nm with an increment of 5 nm. The green dashed line is a guide to eyes indicating

the 3rd order supercavity mode under study. (B) Theoretically obtained scattering peak positions of the modes supported by a single GaAs nanocylinder (H=332 nm) on a substrate in the region of interest of dimensional parameters λ, D. The modes denoted as 3 TE and 3' TE are the ones experiencing the avoided crossing that leads to the formation of the supercavity regime for the 3rd azimuthal order TE mode for which the corresponding Q factors are indicated by red circles (their area being directly proportional to the Q factor value). Mode 4 TE is a higher (4th) order supercavity mode, arising from the anti-crossing with mode 4' TE. Modes 4 TM and 5 TM are regular Mie modes. The green triangles show the experimentally obtained lasing peak positions for cylinders with a fixed height H=330 nm and different diameters. (C) Measured lasing threshold values (shown as red stars) shown as a function of the diameter of the nanocylinder for the supercavity mode 3 TE. The red shadowed area is a guide to eyes. The vicinity of the optimal diameter value is highlighted by the blue dashed line (where the Q factor of the 3 TE supercavity mode has a maximum in theory, ρ=1.47). This has a slight discrepancy with experimental results, where the minimal threshold is obtained at ρ~1.51.

As mentioned above, in order to demonstrate the formation of the supercavity regime of the mode, the lasing threshold is measured and analysed for several nanocylinders with different diameters around the optimum, as shown in Figure 4C. For each diameter, two to four similar nanocylinders are measured to increase the statistic accuracy. As can be seen from the plot in Figure 4C, the lasing threshold has a minimum in the vicinity of D=500 nm, indicating an increase in the Q factor of the mode for this diameter, which is in a good agreement with the theoretically obtained behaviour for the supercavity condition.

In conclusion, using the concept of quasi-bound states in the continuum or supercavity modes, we have demonstrated lasing action at cryogenic temperatures in a single GaAs nanocylinder fabricated on a quartz substrate. The nanolaser has a size-to-wavelength ratio as low as 0.6, which, to the best of our knowledge, is the smallest value demonstrated for purely semiconductor nanolasers so far. This became possible by engineering two sets of resonances supported by the system, namely, Mie resonances and Fabry-Perot resonances, to achieve a quasi-non-radiative mode, the so-called supercavity mode, via mutual destructive interference of the resonances at the targeted lasing wavelength. As compared to previous approaches, the present design does not need the particle to be suspended in air and does not rely on modes of high azimuthal orders, which increases the robustness of the device. We believe that this approach can further be brought to the room temperature operation by using higher gain materials or surface passivation techniques. In the future, with the use of electrical pumping and light out-coupling strategies, the presented concept can pave the way for smaller, Ohmic-loss-free nanolasers, with potential applications in photonic circuits and beyond.

**Methods.**

*Numerical simulations.*

Numerical simulations were performed using an FDTD commercial software (Lumerical FDTD Solutions®). There are two types of numerical simulations carried out in this work: nanoparticle scattering simulations and Q factor calculations. In nanoparticle scattering simulations, the cylinder was irradiated by a plane wave using total-field scattered-field (TFSF) source. This source separates the scattered field from the incident field, which allows to compute scattering efficiency $Q_{sca}$. The scattering efficiency was obtained using the following relation[46]: $Q_{sca}(\lambda)=W_{sca}(\lambda)\cdot(I_s(\lambda)\cdot D\cdot H)^{-1}$ where $\lambda$ is the wavelength of light, $W_{sca}$ is the power scattered through a box monitor (closed surface) surrounding the cylinder, $I_s$ is the intensity of the light source irradiating the nanoparticle, D is the diameter of the cylinder, and H is its height. Two types of plane waves were considered for excitation, namely, TE polarized ($E_z=0$, $H_z\neq 0$, Figure 1A) and TM polarized ($H_z=0$, $E_z\neq 0$), where z is the axis of the cylinder and $E_z$ and $H_z$ are the projections of electric field and magnetic field onto the z axis, correspondingly. From the scattering simulation, the scattering efficiency map $Q_{sca}(\lambda, D)$, shown in Figure 1C, was obtained. To compute the (quasi-normal) eigenmodes, and the corresponding Q factors, one electric dipole was placed inside the dielectric cylinder to selectively excite the modes. To do so, the dipole was placed in the vicinity of the antinode of the electric field of the desired mode with its dipole moment parallel to the electric field at that point (as, for example, depicted in Figure 1B, for excitation of quasi-TE modes). Time-dependent electric and magnetic fields were recorded using 9 point-like time monitors, randomly placed inside the cylinder. After that, the spectrum was calculated for each monitor using Fast Fourier Transform (FFT). Then the spectra were averaged over the number of monitors obtaining the spectrum for Q factor calculations. The resonant peaks in the spectrum were fit by a Lorentzian function to obtain $\Delta f_i$ – FWHM for the $i^{th}$ resonance. The Q factor of the $i^{th}$ resonance was calculated using $Q= f_i/\Delta f_i$ where $f_i$ is the frequency of the $i^{th}$ resonance. The far-field 3D pattern of the modes was computed using a box monitor completely surrounding the cylinder. The far-field 2D pattern of the modes was obtained using a plane monitor (perpendicular to the z axis) located above the cylinder.

In order to account for the real experimental conditions, additional numerical simulations and optimization were performed considering that the dielectric nanocylinder is placed on a quartz substrate and has a capping HSQ layer on top, as can be seen in the SEM image in the inset of Figure 2B. In those simulations, we assumed that the cylinder had a constant relative permittivity $\varepsilon=13$, that the HSQ cylinder on top had a height of 150 nm and a constant refractive index $n_{HSQ}=1.41$ and that the substrate had a constant refractive index $n_{substrate}=1.46$.

*Sample nanofabrication.*

GaAs wafer with an epitaxial structure of i-GaAs (500nm) / AlAs (100 nm) / undoped <100> GaAs (625 nm) was purchased from Semiconductor Wafer Inc., Taiwan. Epitaxial lift-off[37] technique was used to transfer the top, high quality GaAs film (500 nm) from the wafer to a quartz substrate. In the process, black wax (i.e., Apiezon W) was applied on the top i-GaAs layer and thermally annealed at 90°C for about 30-60 min to realize a slightly domed surface. After that, hydrofluoric acid (HF)) solution (~ 5 wt. % in H₂O) was used to etch the sacrificial layer (AlAs) and GaAs epitaxial film was then lifted off. Direct bonding process was used to attach the film to the quartz substrate using a wafer bonder system (EVG 510). Inductively coupled plasma reactive ion etching (ICP RIE) tool (Oxford PlasmaLab System 100) was used to etch GaAs film to a desired thickness. Standard electron beam lithography procedure (Elionix ELS-7000 system) was used to fabricate nanocylinders mask using hydrogen silsesquioxane resist (HSQ, Dow Corning XR-1541-006) along with ESpacer (E-spacer 300Z) to improve the conductivity during e-beam exposure. The unexposed resist was removed using 25% tetramethylammonium hydroxide (TMAH) developer. Finally, the patterned GaAs film was etched using ICP RIE. Nanocylinders were fabricated with a 20 µm spacing. That period was chosen in such a way that each cylinder would be individually irradiated during subsequent optical measurements, thus acting as an isolated nanoparticle.

*Optical characterisation.*

All optical measurements were performed in a micro-spectrometer setup which consists of an inverted microscope (Nikon Ti-U) and a spectrometer system (Andor SR303i spectrograph coupled with a 400 x 1600-pixel Electron Multiplying Charge-Coupled Devices - Newton 971 EMCCD). For lasing and photoluminescence measurements, a Ti:Sapphire femtosecond laser (Coherent®, MIRA 900 with 9450 Reg A amplifier) was used. The pumping wavelength was tuned to ~ 530 nm by using an optical parametric amplifier (Coherent® OPA). The pulse duration and repetition rate after OPA are ~ 200 fs and 20 KHz, respectively. The laser beam was focused on the nanocylinder using a 50X dark-field (DF) microscope objective (Nikon, CFI LU Plan DF, numerical aperture NA = 0.55) resulting in a laser spot of ~ 5 µm in diameter. The dark-field scattering was measured in a backward scattering configuration using the same microscope objective with white light excitation from a halogen lamp. Transmission of GaAs thin film was measured using another halogen lamp excitation. The absorption was calculated by the formula A = 2 − $\log_{10}$(%T). All spectrally resolved measurements were carried out with 150 groves/mm, 800 nm blading grating and the spectrograph slit width of 100 µm resulting in a spectral resolution of ~ 1 nm. All measurements were performed at cryogenic temperatures. For that, the sample was placed onto the cold-finger of a cryostat (Janis, ST-500). The chamber was vacuumed to 3x10$^{-3}$ Torr by an oil pump before being cooled down to 77 K using liquid nitrogen.

*Auto correlation measurements.*

Auto correlation measurements were carried out using Hanbury–Brown–Twiss set-up. The emission signal in our microscopic lasing measurement setup was routed into a multimode fiber (Thorlabs GIF625, 62.5 μm core). The signal was then sent to a 50:50 optical fiber beamsplitter with each node coupled with a low-jitter avalanche photodiodes (Micro Photon Devices, PDM series with a timing jitter of 35ps FWHM). Coincident clicks on the two photodiodes was recorded using a time-tagging device (qutools, QuTAU, timing resolution 81ps).

**Author Contributions**

R.P.-D. and A.I.K. conceived the idea. V.M. performed numerical simulations. Z.P. performed epitaxial lift-off of the GaAs film, its bonding to the quartz substrate and etching. V.M. performed etching, electron beam lithography and etching after EBL. V.V. performed SEM measurements. S.T.H. constructed the optical setup, performed all optical measurements and their data processing. V.M., S.T.H., R.P.-D., H.V.D. and A.I.K discussed the results. V.M. wrote the first draft of the manuscript and all authors worked on it. R.P.-D., H.V.D. and A.I.K. coordinated and supervised the work.

**Notes**

The authors declare no competing financial interest.

**Funding Sources**

A*STAR SERC Pharos program, Grant No. 152 73 00025 (Singapore).

**Acknowledgments**

The authors acknowledge Leonid Krivitsky (IMRE, A*STAR) and Victor Leong Xu Heng (IMRE, A*STAR) for processing autocorrelation function measurements data.

# Supplementary information

# Lasing action in single subwavelength particles supporting supercavity modes

Vasilii Mylnikov[1,2], Son Tung Ha[1], Zhenying Pan[1], Vytautas Valuckas[1], Ramón Paniagua-Domínguez[1], Hilmi Volkan Demir[2,3], Arseniy I. Kuznetsov[1#]

[1] Institute of Materials Research and Engineering, A*STAR (Agency for Science, Technology and Research), 2 Fusionopolis Way, #08-03, Innovis 138634, Singapore

[2] LUMINOUS! Center of Excellence for Semiconductor Lighting and Displays, The Photonics Institute, School of Electrical and Electronic Engineering, Nanyang Technological University, 50 Nanyang Avenue, Singapore 639798

[3] Bilkent University UNAM - Institute of Nanotechnology and Materials Science, Department of Electrical and Electronic Engineering, Department of Physics, Bilkent University, Ankara, Turkey 06800

# Corresponding author, email: arseniy_kuznetsov@imre.a-star.edu.sg

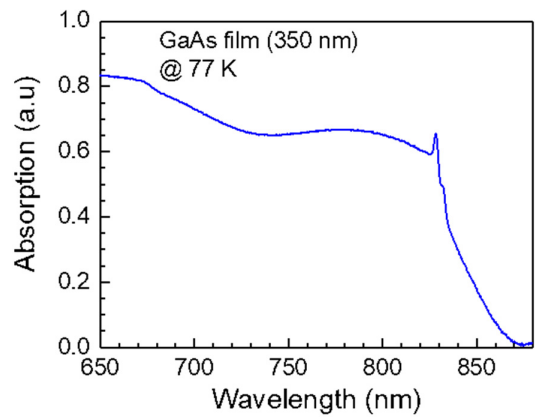

**Figure S1.** Measured absorption of the GaAs film of 350 nm in thickness at 77K.

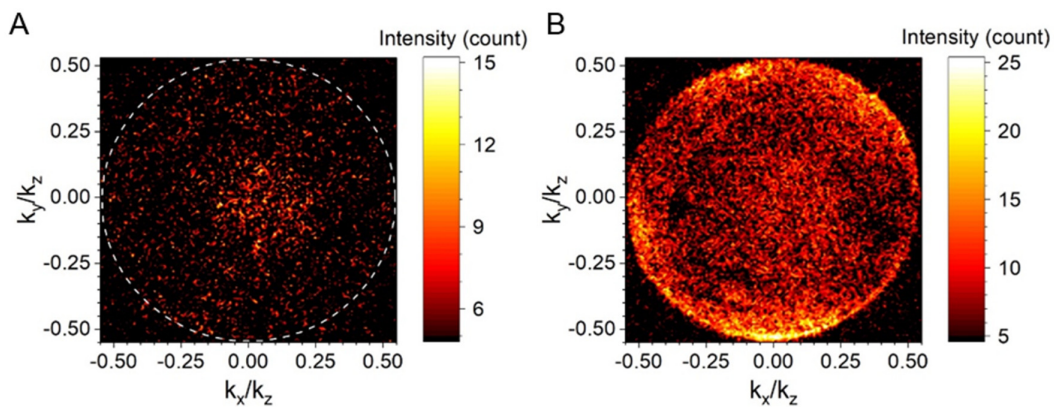

**Figure S2.** Back-focal-plane image of the emission below (A) and above (B) lasing threshold, measured from the GaAs cylinder of D=500 nm and H=330 nm using 0.55 NA objective. Radiation pattern of the pillar in the x-y plane was measured by projecting the back-focal-plane of the collecting objective onto the EMCCD detector (see Methods). It is noted that the NA of the objective used in our setup is only 0.55, thus it can only capture the edge of the side lobe emission of the supercavity mode (see panel B and Figure 1G from the main text). Note that, due to the degeneracy in azimuthal angle (and the random nature of the emission process), the lobes seen in Figure 1G in the main text turn into "rings" in the experimental back focal plane images obtained for lasing emission. We note that, as can be seen from panel B, there are several random brighter spots on the ring pattern, which may come from the different scattering efficiency on the edge of the cylinder, due to fabrication defects.

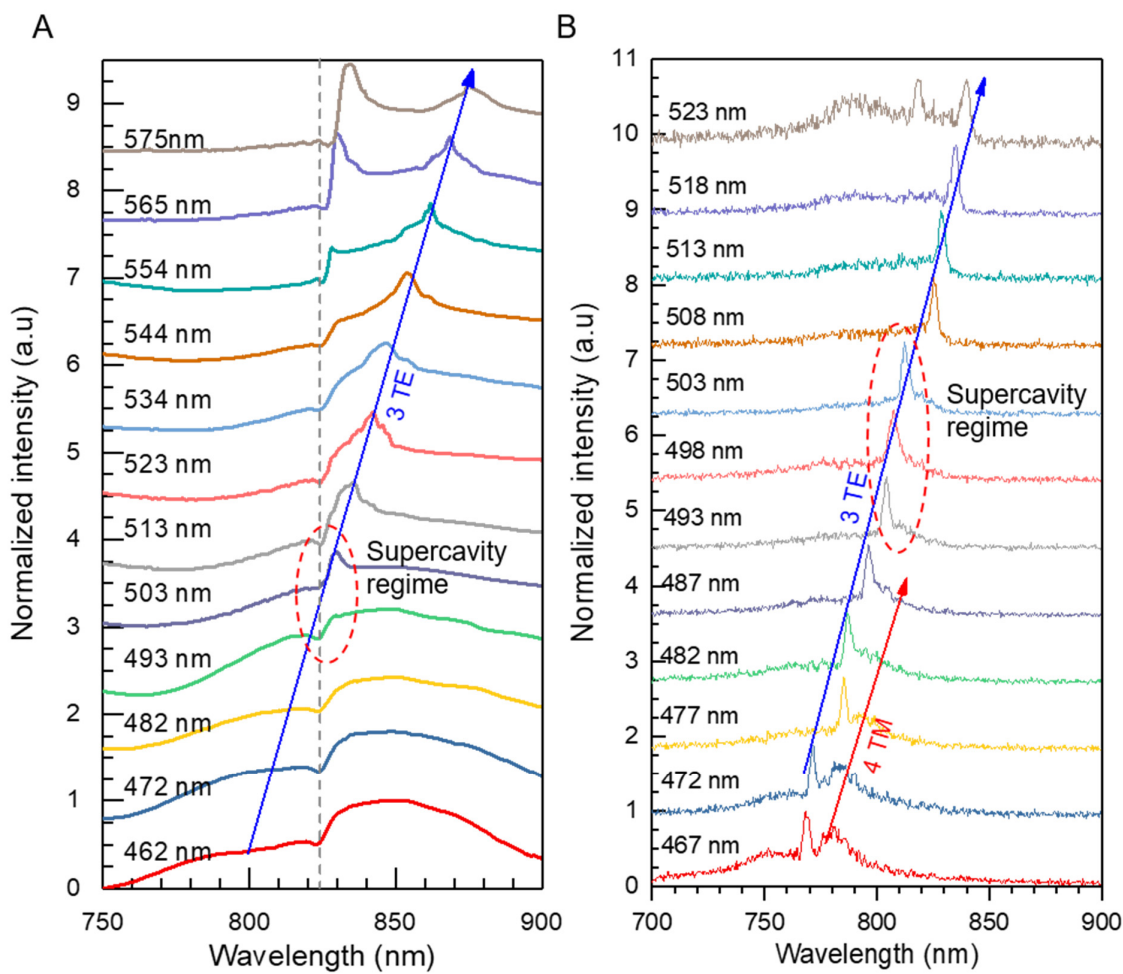

**Figure S3.** Optical characterisation of the GaAs nanocylinders with fixed height H=330 nm and a span in diameters. (A) Normalized experimental dark-field scattering spectra. Dashed line indicates the 3rd order supercavity mode. (B) Evolution of the lasing spectrum for GaAs nanocylinders. The pumping, for each diameter in the plot, was set at a different fluence but always above the lasing threshold. Supercavity mode 3 TE and Mie mode 4 TM identified by numerical simulations are highlighted by arrows.

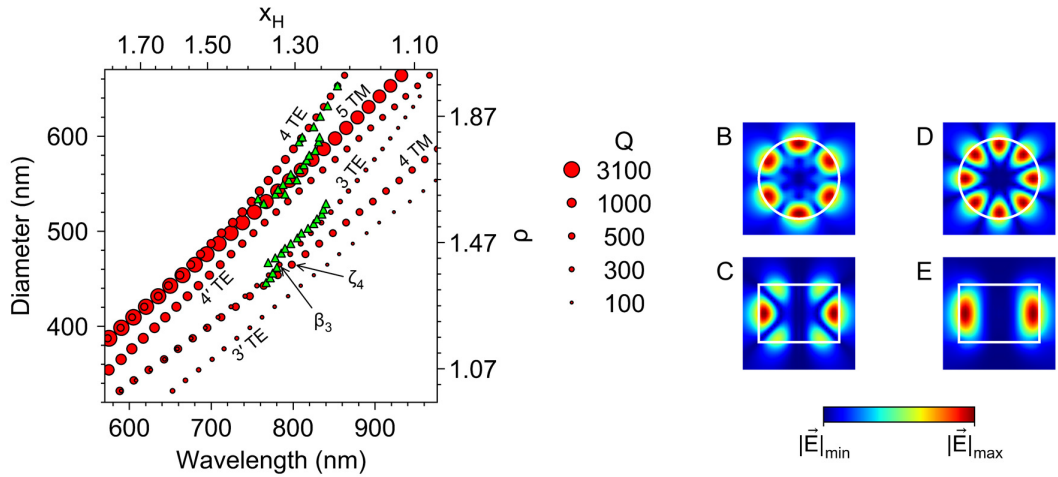

**Figure S4.** Simulated characteristics of the single dielectric nanocylinder (H=332 nm, ε=13) on a substrate ($n_{substrate}$=1.46) with HSQ cap on top ($H_{HSQ}$=150 nm $n_{HSQ}$=1.41). (A) Theoretically obtained scattering peak positions of the modes supported by a single dielectric nanocylinder. Q factors of the modes are shown by red circles, where the area of the circles is directly proportional to the value of the Q factor. Modes 3 TE and 3' TE experience avoided crossing (mode 3 TE is the supercavity mode). Mode 4 TE is the 4th order supercavity mode which anti-crosses with the mode 4' TE. Modes 4 TM and 5 TM are regular Mie modes. The above-mentioned modes were added to the graph because they lay in the region of interest of the dimensional parameters λ, D. The green triangles show lasing peak positions obtained from experiments for the cylinders with a fixed height H=330 nm and varied diameters. Q=270 is the Q factor of the mode 3 TE at the point $β_3$ (D=465 nm, λ=783.5 nm). Q=600 is the Q factor of the mode 4 TM at the point $ζ_4$ (D=465 nm, λ=798.3 nm). In spite of the fact that Q factor of the mode 4 TM is nearly 2 times higher than that of the supercavity mode 3 TE at D=465 nm, when the diameter exceeds 460 nm, the lasing is favourable to happen in the 3rd supercavity mode instead of 4th order Mie mode even though both of them are lying within the gain spectrum. This stems from the fact that resonances of higher azimuthal orders are more sensitive to the sidewall roughness created during the dry-etch process, which tends to lower the Q factor in experiments. In this regard, the overlap of the near field maxima with the sidewalls of the cylinder is more pronounced for the mode 4 TM compared to that of the mode 3 TE. (B, C) Simulated near-fields of the supercavity mode 3 TE for the parameters corresponding to the point $β_3$ in XY (B) and YZ (C) planes of the cylinder. (D, E) Simulated near-fields of Mie mode 4 TM for the parameters corresponding to the point $ζ_4$ in XY (D) and YZ (E) planes of the cylinder. White lines in (B-E) show cylinder boundaries. Consequently, despite having a lower theoretical Q factor, the mode 3 TE seems to be more robust against fabrication imperfections and therefore more suitable to obtain lasing in practical devices.